\documentclass[iop]{emulateapj}
\usepackage[version=3]{mhchem}
\usepackage{wasysym}

\newcommand{\LampeEmail}{lampe@clermont.in2p3.fr}

\newcommand{\GS}{GS~1826$-$24}
\newcommand{\KS}{KS~1731$-$26}
\newcommand{\UHURU}{4U~1636$-$536}

\newcommand{\MXB}{MXB~1730$-$335}

\newcommand{\TFIVE}{IGR~J17480$-$2446}

\slugcomment{Draft of \today; v.1.0} 

\shorttitle{Trends in X-ray burst models}

\shortauthors{Lampe, Heger \& Galloway}

\begin{document}


\title{The influence of accretion rate and metallicity on thermonuclear bursts: predictions from KEPLER models}


\author{Nathanael Lampe\altaffilmark{1}, Alexander Heger \& 
Duncan K. Galloway}
\affil{School of Physics and Astronomy, Monash University, VIC 3800, Australia}


\email{\LampeEmail}

\altaffiltext{1}{Current address: Universit\'{e} Clermont Auvergne, Universit\'{e} Blaise Pascal, CNRS/IN2P3, Laboratoire de Physique Corpusculaire, BP 10448, F-63000, Clermont-Ferrand, France}


\begin{abstract} 

Using the {\sc KEPLER} hydrodynamics code, 464 models of thermonuclear X-ray
bursters were performed across a range of accretion rates and compositions. We
present the library of simulated burst profiles from this sample, and examine variations
in the simulated lightcurve for different model conditions. We find that the recurrence 
time varies as a power law against accretion rate, and measure its slope while mixed H/He 
burning is occurring for a range of metallicities, finding the power law gradient to vary from $\eta = 1.1$ to $1.24$. We also identify the accretion rates at
which mixed H/He burning stops and a transition occurs to different burning regimes. We
also explore how varying the accretion rate and metallicity affects burst
morphology in both the rise and tail.

\end{abstract}

\section{Introduction}

Thermonuclear (type I) X-ray bursts occur in low mass X-ray binaries when the
fuel accumulated onto the neutron star from it's donor undergoes
 unstable nuclear burning \citep[for reviews see
][]{Lewin1993Review,Strohmayer2006Review}.
Whilst burst sources exhibit significant
variation in burst properties with changes in accretion rate and spectral state,
large databases of burst observations \citep[such as][]{Galloway2008} can be
used to identify trends and patterns of variation common to all sources,
or sub-groups with common properties.
This is complemented by a growing ensemble of X-ray burst models, which permit: 
observation of the expected effects of varied atmospheric metallicities on both nuclear processes and
observable burst properties  \citep{Jose2010MetallicityModels}; comparisons of
models with observations \citep{Heger2007}; simulations of superbursts
\citep{Keek2011Superbursts}; and simulations of quasi-periodic oscillations in
burst sources \citep{Heger2007QPO}. Burst
modelling codes vary in complexity from simulating energy generation in a single zone
with a limited reaction network \citep[e.g., ][]{Taam1980Models,
Hanawa1982ModelsHHe, Cumming2000ModelsXray}, whilst more complex models
simulate hydrodynamical and nuclear processes and track chemical abundancies in
the neutron star atmosphere across multiple zones \citep[e.g.,
][]{WallaceWoosley1982KEPLER,Fisker2008Bursts}.

Models of thermonuclear bursts predict variations in burst
behaviour that arise due to the changing conditions of the accreting neutron
star. In particular, bursts show substantial variation in behaviour with
accretion rate \citep{Fujimoto1981accretion,Narayan2003StabilityRegimes} that
affects their morphology (Table \ref{tab:regimes}). At high accretion rates the
accreted fuel burns stably (case I), whilst at lower accretion rates bursts with a
long tail from \emph{rp}-process \citep{Wallace1981ExplosiveHydrogenBurning} burning occur (case II/III). At still lower accretion rates
the hydrogen burns to helium in the lower layers of the atmosphere (case IV). Ignition
of the helium layers in these pure He-fuelled bursts leads to bursts that typically exhibit photospheric radius expansion (PRE), and
correspondingly shorter burst durations. Within these broad categories of
bursting behaviour, changes in accretion rate
\citep[e.g.,][]{Thompson2008recurrenceTime}, rotation speed 
\citep{Linares2012Terzan5,Bagnoli2013SlowRotator}, atmospheric composition 
and surface gravity
\citep[e.g.,][]{Suleimanov2012AtmoModels} and the location of burst ignition
\citep[e.g.,][]{MaurerWatts2008IgnitionLatitudeAndConvexity} are expected to drive subtle
variations in the morphology, spectra and regularity of bursts.

Despite the variations in burst morphology within most sources, some
neutron stars such as \GS\ \citep{Ubertini1999Clocked} and \KS\
\citep{GallowayLampe2012Consistency} have been noted for showing consistent
burst profiles. Additionally, \GS\ shows remarkable consistency in its
recurrence time. As a result, these sources have been frequently used to
test current understanding of burst theory. Given that in burst models we
simplify the neutron star to a one-dimensional column with a simplified
atmosphere and no accretrion disk, one would expect a similar regularity in
modelled lightcurves to these well studied and regular sources. Interestingly however,
the consistency of burst models in comparison to the consistency of known regular sources has not been widely explored.
%
%

Whilst models are successful in reproducing burst energetics, timescales and
rates \citep[e.g.,][]{Fujimoto1981accretion,Cumming2000ModelsXray} for certain
systems, such as pure He bursts in SAX~J1808.4$-$3658 \citep{Galloway2006SAX}
and mixed H/He bursts in EXO~0748$-$676 \citep{Boirin2007Triplets}, theoretical
predictions of burst behaviour are not always matched by observation.
\cite{Cornelisse2003BeppoSAX} found several burst sources that show the burst
rate decreasing with increasing persistent flux, contrary to expectations that
increased accretion increases the frequency of bursts \citep[cf. with][]{Galloway2008}.

Occasionally type I X-ray burst sources are known to exhibit two distinct peaks
in the burst rise. This can occur in PRE bursts where cooling of the photosphere
causes spectral changes that shift the emission out of the X-ray band, yet maintaining
a single peak in bolometric flux
\citep{Paczynski1983HeModels}. More interestingly, twin peaks have
been seen in the bolometric flux from a number of sources, such as: 4U~1608$-$52
\citep{Penninx1989DoublePeak1608}; \UHURU\
\citep[e.g.][]{Stazjno1985DoublePeak1636,Galloway2008}; XTE~J1709$-$267
\citep{Jonker2004DoublePeak1709}; and GX~17+2
\citep{Kuulkers2002DoublePeakGX17}. Various authors have proposed models for 
these structures including 
modulation of the bolometric flux by a burst induced accretion disk corona 
\citep{Melia1992DoublePeaks},
variations in ignition location towards the poles of the neutron star causing a
stalling of the burning front as it approaches the equator
\citep{Bhattacharyya2006DoublePeaksPropagation,WattsMaurer2007TwinPeaksAccRate},
or a multi-stepped energy release, caused by either a waiting point in the
$rp\mathrm{-process}$ burning that stalls the reaction rate
\citep{Fisker2004WaitingPoints}, or by hydrodynamic instabilities that lead to
mixing \citep{Fujimoto1988TwoPeaks1636}. Given consistent physical parameters such a
morphology could be expected to arise more regularly than is seen in observation,
though \cite{Jose2010MetallicityModels} do find sparsely occuring double peaks
within burst simulations that could be linked to varying burst compositions.
Adding to this irregularity is the observation of triple-peaked structures in
\UHURU\ \citep{vanParadijs1986TripleBurst,Zhang2007TripleBurst}.


In this study we analyse simulations of thermonuclear bursts made using the 1-D
hydrodynamics code, {\sc KEPLER} \citep{Woosley2004}. In Section \ref{sec:method} we
describe our method for analysing the simulated lightcurves. We present the
results and discussion of the model properties in two sections. Section
\ref{sec:rectime} looks at how variation of the initial model conditions impacts
the burst recurrence time and the bursting regime. Variations in the morphology
of the burst rise and tail, as well as the appearance of twin-peaked bursts from
nuclear processes are considered in Section \ref{sec:morphology}.

\begin{table}[]
\centering
\setlength\tabcolsep{2pt}
\begin{tabular}{cccc}

Case & $L_{\mathrm{acc}}/L_{\mathrm{Edd}}$ & Steady Burning & Burst Fuel \\
\hline
\hline
I  & $\ge 0.25$     &  stable H/He     & none       \\
II & $0.15 - 0.25$ & overstable H/He & mixed H/He \\
III& $0.04-0.15$   &  stable H        & mixed H/He \\
IV& $0.004-0.04$ &  stable H        & pure He    \\
V & $\le 0.004$    &  none            & mixed H/He \\
\setlength\tabcolsep{5pt} 
\end{tabular}

%
%
\caption{Burning regimes described by \cite{Narayan2003StabilityRegimes} using a
one-dimensional linear stability analysis for a $1.4\,\mathrm{M_\odot}$ neutron
star with $R=10.4\,\mathrm{km}$ and core temperature $T=10^{7.5}\,\mathrm{K}$.
\label{tab:regimes}}

\end{table}

\section{Analysis}\label{sec:method}
We simulated thermonuclear bursts for a range of initial conditions using the
1-D implicit hydrodynamics code {\sc KEPLER} \citep{Woosley2004}.  {\sc KEPLER} is a multizone model that
tracks elemental abundancies at different depths in the stellar atmosphere. It
was designed to model explosive astrophysical phenomena
\citep{weaver1978KEPLER}, and currently models a thermonuclear reaction network consisting of over  1,300 isotopes in addition to
convective mixing and accretion.

{\sc KEPLER} was designed to model both hydrostatic and explosive astrophysical phenomena (Weaver et al. 1978).  We use an adaptive thermonuclear reaction network that adds species and the corresponding nuclear reactions as needed (abundances rise above some very conservative threshold) out of a data base of over 6,000 nuclei from hydrogen to polonium and up to the proton drip-line, well covering the nuclei relevant to $\alpha$p and rp process.  Convection is modelled using time dependent mixing-length theory, and we also include semiconvection \citep{weaver1978KEPLER} and thermohaline mixing \citep{Heger2005Presupernova}.  Composition mixing is modelled as a (turbulent) diffusive process with the diffusion coefficients derived from the respective processes above.  The models are based on non-rotating neutron stars using a local Newtonian frame. Accretion is modelled by allowing the pressure at the outermost zone to increase in order to simulate the weight of the accreted material above it.  Once
the pressure exceeds the value that corresponds to the minimum mass of a new zone, a new zone is added to the simulation containing the accretion material and it's outer pressure is set to a value corresponding to the remainder of the accreted material, allowing for continuous accretion process.  A models may accumulate many thousands of zones small zones that that are de-refined as they move deeper inside the star.  In many runs, the thinnest zones at the bottom of the accreted material have a spatial resolution of less than 1 cm.

For the metals ($\mathrm{Z}$) in the accretion composition, we use just $^{14}$N for simplicity, consistency, and for historic reasons \citep{Taam1993SuccessiveBursts}.  This is also the dominant metal one would expect from an evolved companion star in which the material has undergone some CNO processing.  Once accreted, the material very quickly reaches the beta-limed (hot) CNO cycle with equilibrium abundances of $^{14}$O and $^{15}$O independent of initial relative distribution CNO isotopes\footnote{There are, however, small changes in the total molarity of CNO isotopes for a given value of $Z$ depending on relative isotope distribution.  We did not include other metal in the accreted material; the deeper layers are quickly dominated by XRB ashes and the opacity in most of the other layers is not affected much.}. Additional details of the simulation are described in \cite{Woosley2004}.

In this analysis, we use the modeled
light curves taken at the outermost edge of the photosphere to quantify the
behaviour of the models, calculating equivalent parameters to those used in
observational studies. The outermost edge of the photosphere in {\sc KEPLER} models
represents the surface of last scattering and is closest to what would
be seen observationally. The radius of this zone is also tracked during the simulation,
providing a measure of atmospheric dynamics, however analysis of it is beyond the scope of this paper. The 
reaction rates chosen as an input to {\sc KEPLER} are identical to those in \cite{Woosley2004}.

We ran simulations for a range of metallicities ($Z$) and accretion rates
(Table \ref{param-space}), where the accretion rate
$l_\mathrm{acc}=L_\mathrm{acc}/L_\mathrm{Edd}$ is given as a fraction of the
Eddington accretion luminosity,
$L_\mathrm{Edd}=2.05\times10^{38}\,\mathrm{erg}\,\mathrm{s^{-1}}$. The
accretion rate in $\mathrm{M_\odot}\,\mathrm{yr^{-1}}$ is found by multiplying
$l_\mathrm{acc}$ by the Eddington rate
$\mathrm{\dot{M}_{Edd}}=1.75\times10^{-8}\,\mathrm{M_\odot}\,\mathrm{yr^{-1}}$.
Where the accretion rate and composition is appropriate for bursts to occur, a sequence of bursts or `burst train' is said to occur. Burst trains occurred at most accretion rates, switching to steady
burning above a critical accretion rate. We have not explored variation of the
hydrogen fraction (H) of the accretion rate independently of metallicity, rather
we assume the typical helium fraction (He) varies with metallicity according to $\mathrm{He} \approx 0.24 + 1.75\mathrm{Z}$, and from this the Hydrogen fraction can be found as $\mathrm{H} = 1-\mathrm{He}-\mathrm{Z}$.


We analysed each model run to identify individual bursts, by searching for
local maxima that follow an increase in the modelled luminosity by a factor of
$\ge10$, compared to the luminosity $20\,\mathrm{s}$ prior to the maxima. The
luminosity $20\ \mathrm{s}$ prior to the burst is adopted as the persistent
thermal luminosity. In most models this criterion is sufficient to identify the
beginning of a burst, however some models exhibit convective shocks reaching the surface which appear
as local maxima preceeding the peak of the burst \citep{2012KeekSuperburstModels}.
These shocks typically last less than $100\,\mathrm{ms}$ and are associated
with pure He bursts or mixed H/He bursts that have depleted a substantial amount
of their accumulated hydrogen by hot CNO burning prior to ignition. In these
cases, rapid convection following a helium flash leads to sudden heating of the
outermost zone of the neutron star, triggering a momentary surge in the
luminosity of the atmosphere. These events are a local phenomenon which would 
not be expected to be detectable in observations. On
a neutron star the surge in luminosity would be isolated to a single convective
cell, and would be negligible compared to the collective emission from every such cell 
on the star. In the models, by assuming a single column represents the entire 
stellar surface, we predict a rapid increase in brightness across the entire star.
To avoid interpreting shocks as burst peaks, we screen each maximum to ensure that 
it is not a shock, but rather the lightcurve maximum.

In the most severe cases, rapid convection from a helium flash can bring enough
hot material to the surface to increase the luminosity above
$10^{3}L_\mathrm{Edd}$ over a window of $20\,\mathrm{ms}$. To make these models
more amenable to analysis, models that exceed
$10^{39}\,\mathrm{erg}\,\mathrm{s^{-1}}$ (the total luminosity being found by
integrating the simulated column over the neutron star surface) are rebinned
into uniform $125\,\mathrm{ms}$ timesteps, roughly equivalent to the
time-binning in many observational studies \citep{Galloway2008}.

Not every simulation produces a train of bursts. A considerable number of
simulations were run at accretion rates close to the transition to stable
burning. These models often show one burst followed by stable or quasi-stable
burning for extended periods, accompanied by elevated luminosities.

\begin{table}[]
\centering
\begin{tabular}{ccccc}

$Z$ & $H$ & $l_\mathrm{acc}$ & \multicolumn{2}{c}{Models} \\

\hline \hline

0.000 & 0.7600 & 0.010--50 & 49 & (30)\\
0.001 & 0.7590 & 0.010--50 & 54 & (35)\\
0.002 & 0.7545 & 0.010--50 & 55 & (39)\\
0.004 & 0.7490 & 0.010--50 & 50 & (31)\\
0.010 & 0.7324 & 0.010--50 & 50 & (34)\\
0.020 & 0.7048 & 0.002--50 & 63 & (42)\\
0.040 & 0.6496 & 0.010--50 & 59 & (40)\\
0.100 & 0.4840 & 0.010--50 & 40 & (25)\\
0.200 & 0.2080 & 0.010--50 & 37 & (24)\\


\end{tabular}

\caption{Metallicities, hydrogen fractions and the accretion rates spanned for
the set of {\sc KEPLER} burst models analysed in this report. The number of model runs
at each composition is given, with bracketed values showing the number of runs
with four or more bursts.\label{param-space}}

\end{table}

We analysed each burst from $20\,\mathrm{s}$ prior to the peak through to the
end of the burst. The luminosities neglect the persistent contribution
from accretion, so we adopt the value of the emission $20\,\mathrm{s}$ prior to
the peak luminosity as the thermal persistent emission, $L_\mathrm{th}$ (as it
arises from the thermal properties of the neutron star atmosphere). We define
two criteria to mark the end of a burst. First, a drop in the luminosity to
$L=L_{\mathrm{th}}+0.02(L_{p}-L_{\mathrm{th}})$, where $L_{p}$ is the peak burst luminosity, may mark the end of a burst,
however at high accretion rates the burst tail may become extended and show
resurgences in luminosity from overstable burning. Hence, from $50\,\mathrm{s}$
after the burst peak, if the burst tail rises in luminosity above what it was
$10\,\mathrm{s}$ prior the burst end is also triggered. For all bursts, we
insist that the burst extends at least $10\,\mathrm{s}$ beyond the peak luminosity.
Any deviations from this analysis were flagged according to the scheme described in
Appendix \ref{app:AnalyisisFlags}.

\subsection{Burst parameters}
We calculated parameters describing each burst in the train, which were
averaged to give characteristic values for each set of initial
conditions. We briefly cover the definition of each parameter in the following
paragraphs.


The burst fluence, $E_b$, is determined by numerically integrating the burst luminosity with time
\begin{equation}
E_{b}=\int_{t_{\mathrm{start}}}^{t_{\mathrm{end}}}\left(L(t)-L_{\mathrm{th}}\right)\ dt,
\end{equation}
where the persistent thermal emission is subtracted from the burst prior
to integration. Whilst $t_{\mathrm{start}}$ may be well before the burst rise,
luminosities here contribute negligibly to the burst fluence as they are close
to the thermal level before the burst begins. We calculated the ratio of fluence to peak
flux ($\tau=E_{b}/L_{p}$) as it is often used as a measure of burst
morphologies \citep{VanParadijs1988Tau}.

We also consider the ratio of persistent fluence between bursts to the burst
fluence, commonly referred to as $\alpha$ \citep{Gottwald1986Alpha}. The alpha
ratio is typically found by integrating the persistent emission, however as
accretion is constant within the models studied and the dominant source of
persistent emission, we calculate $\alpha$ as follows
\begin{equation}
\alpha = \frac{E_p}{E_b} = \frac{L_\mathrm{acc}+L_\mathrm{th}}{E_b}\Delta t
\end{equation}
where $\Delta t$ is the time since the previous burst and $L_\mathrm{acc}$ is
found from the accretion rate. Note that this definition restricts us from
calculating the value of $\alpha$ for the first burst in a train as the
recurrence time since the last burst is unknown. In the Newtonian 
frame $L_\mathrm{acc}$ is
\begin{equation}
L_\mathrm{acc} = \frac{G{M}\dot{{M}}}{r}.
\end{equation}
This neglects the gravitational redshift under 
which requires a correction for a distant observer (described in
Appendix \ref{app:redshifts}). Note that, the accretion luminosity exceeds
the thermal persistent luminosity by at least an order of magnitude (typically two orders of magnitude).

To quantify the burst rise both its duration and morphology are
measured. The duration is measured by the time taken to reach 10\%,
25\% and 90\% of the peak luminosity above the thermal level, given by $t_{10}$,
$t_{25}$ and $t_{90}$ respectively (we additionally define the overall burst duration as
the time from $t_{25}$ to the end of the burst). The rise times are then given
in two ways (to enable comparison with existing measures), by
$t_{10-90}$ and $t_{25-90}$, the times respectively taken to rise from 10\%
and 25\% of the peak flux to 90\% of the peak flux.
The shape of the rise can be measured by the convexity ($\mathcal{C}$) of the
burst. \cite{MaurerWatts2008IgnitionLatitudeAndConvexity} define the burst convexity as
the deviation of the lightcurve from linear during the rise from $t_{10}$ to
$t_{90}$. Observationally, variations in convexity can be correlated with the
latitude of burst ignition on a neutron star. The convexity is calculated by
rescaling the burst between $t_{10}$ and $t_{90}$ in both time and luminosity so
each parameter varies from 0 to 10. Calling the rescaled luminosity $l$, and the
rescaled time $x$, the line joining the two parameters are equal along the line
$l=x$. The lightcurve is then a distance of $l-x$ above the straight line
joining the two boundary luminosity points. Integrating this difference, the
convexity can be found as
\begin{equation}
\mathcal{C} = \int_{0}^{10}\left(l-x\right)\ dx. 
\end{equation}
We evaluate this integral numerically for each burst profile. Additionally,
as the models do not have a uniform time resolution, we approximate $t_{10}$ and
$t_{90}$ by linearly interpolating between the two points closest to
$L=0.1L_{p}$ and $L=0.9L_{p}$.

Exponential fits are frequently made to the cooling tail of bursts
\citep[e.g.][]{Galloway2004}. Recently, \citep{Intzand2014CoolingTails} suggested that
the burst tail is better fit by a power law. We fitted single
tailed exponential curves and power laws to each lightcurve. The
exponential fit is given by
\begin{equation}
L(t) = L_{0}\exp \left(-(t-t_{0})/t_{b}\right) + L_{\mathrm{bg}},
\end{equation}
whilst the power law is fitted by
\begin{equation}
L(t) = L_{0}\left(\frac{t-t_{s}}{t_{0}-t_{s}}\right)^{-\kappa} + L_{\mathrm{bg}}.
\end{equation}
In both of these cases the persistent luminosity is measured as the luminosity
$20\,\mathrm{s}$ before the burst peak, and $t_{0}$ is the time of the first
fitted data point. In the case of the exponential fit, this leaves two free
parameters, $L_{0}$ and $t_{b}$, whilst the power law has three free parameters,
$L_{0}$, $t_{s}$ and $\kappa$. The most noticeable difference between the two
curves is the divergence of the power law near $t_{s}$, a time loosely
corresponding to the burst start. Due to the nature of the power law, $t_{s}$
and $\kappa$ are strongly coupled.

To determine the range of the fit, \citet{Intzand2014CoolingTails} varied $t_{0}$, increasing this
parameter's value until a local minimum in $\chi^2_\nu$ is reached. We instead fitted the
burst tail, from the time where the luminosity drops to 90\% of its peak value to
the time where it falls to 10\% of its peak value above the persistent thermal
level. As each separated lightcurve does not have any error estimates, we fit via
a least-squares minimisation.

\subsection{Averaged lightcurves}
The individual bursts are also averaged together to produce mean luminosity and
radius profiles for each model. This provides a single representative burst for
each run. Where three or more bursts occur, the first burst is ignored,
as it is typically substantially different to following bursts, as it lacks an
ashes layer from previous bursts \citep{Woosley2004}. In models that produce only two bursts the first two bursts are averaged.  Whilst this leads to models that have only two bursts showing a significantly higher variation in their parameters than those with three or more, on account of the different fuel column in the first burst, keeping this burst in the averaging process yields a measure of luminosity variation that would be otherwise lost.
The actual luminosities that are
averaged use the duration of the longest burst found for the train, rather than
averaging bursts of different durations. During the averaging process, bursts are aligned so that the times of their peak luminosities coincide.

We present the averaged lightcurves and their data files
online\footnote{http://burst.sci.monash.edu/kepler}, to accompany the averaged burst parameters. The
averaged lightcurves provide luminosities and the 1-$\sigma$ variation in luminosity. The averaged radius
of the last zone in KEPLER is also provided for each model with it's 1-$\sigma$ variation.

\subsection{Analysis of observational data}
We also compare observations of \GS\ to model predictions within this paper,
using bursts observed by \emph{RXTE} \citep{RXTE1996Calibration} in the
MINBAR\footnote{MINBAR is the Multi INstrument Burst ARchive, and can be found at
http://burst.sci.monash.edu/minbar} burst archive. The data analysis techniques used to
produce MINBAR are similar to those found in \cite{Galloway2008}.
\emph{RXTE}  provides time resolved spectra across
the energy range $2-60\ \mathrm{keV}$, which, during each burst, were binned in
$0.25\ \mathrm{s}$ intervals during the burst rise and peak. During the tail, the 
time-bins were increased in order to preserve the signal-to-noise ratio. The burst
background was estimated from a $16\ \mathrm{s}$ interval prior to the burst.

We accounted for the `deadtime' experienced between photons within the \emph{RXTE}
PCA detectors, which reduces the count rate below that experienced by the detector
(by approximately 3\% at an incident event rate of
$\approx 400\ \mathrm{counts}\ \mathrm{s}^{-1}\ \mathrm{PCA}^{-1}$). Each spectrum was
corrected for deadtime by taking an effective exposure that accounts for the deadtime
fraction. The spectra were subsequently re-fitted with a blackbody model over the interval
$2.5-20\ \mathrm{keV}$ using {\sc XSpec} version 12 \citep{XSpec1996FirstTenYears}
with Churazov weighting to accomodate 
spectral bins with low count rates. The effects of interstellar absorption were modelled 
using a multiplicative component ({\tt wabs} in {\sc XSpec}) with the column density fixed
at $nH=0.4$ \citep{IntZand1999nHValues}.

\section{Results and Discussion I: Variations with recurrence time}\label{sec:rectime}
The burst recurrence time is a fundamental observable for determining burst
ignition conditions. A consistent burst recurrence time indicates consistent
ignition conditions within the neutron star. The largest contribution to
variations in recurrence time comes from changes in the accretion rate, as this quantity
alters the time it takes to accrete a critical column of fuel. Multi-dimensional
effects that we do not simulate such as variations in accretion location due to
accretion disk dynamics, as well as unsteady accretion, can have a large impact
on recurrence time, despite the underlying regularity of the nuclear reactions
in the stellar atmosphere. Perhaps, as a result of this, very few sources show consistent
recurrence times, the most notable regular bursters being \GS\
 and \KS.
Within our simulations, recurrence time is found to be consistent for each model
run, as new bursts require the collection of roughly the same amount of fuel for a
burst to begin.

In this section we explore the variation in recurrence time with accretion rate
for prompt mixed H/He bursts. Outside this case we observe transitions to pure He bursts at
lower accretion rates and to delayed mixed H/He bursts at higher accretion
rates. We explore the variation of effective burst duration
($\tau$) with recurrence time observationally in \GS, and compare this to the
loci traced by {\sc KEPLER} models through the $\Delta t-\tau$ parameter space as
accretion rate changes.

\subsection{Variation with accretion rate}
\begin{figure*}
\centering
\includegraphics[]{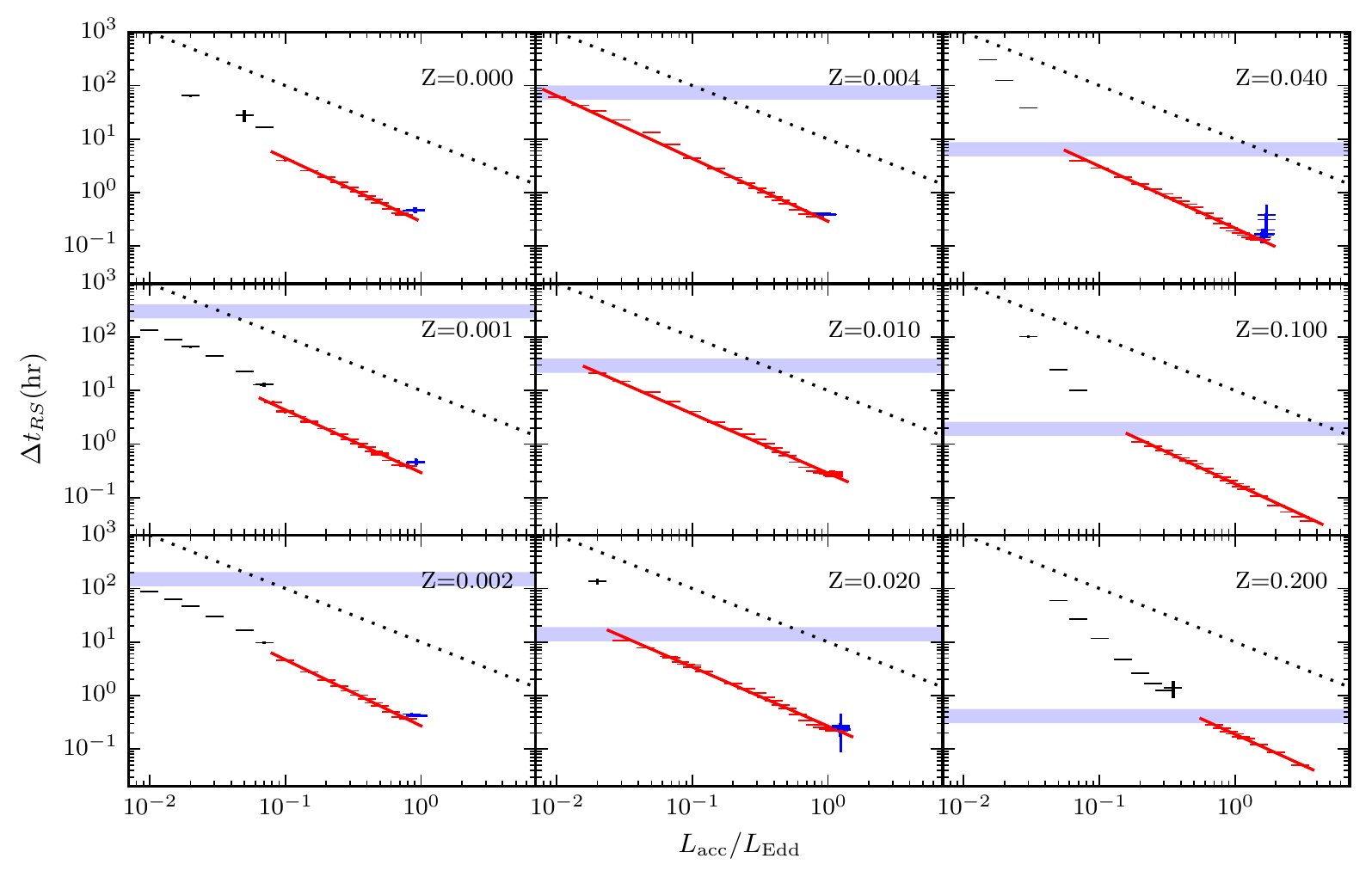}
\caption{Burst recurrence time ($\Delta t_{RS}$) as a function of accretion rate for different 
accreted metallicities, redshifted by $z=1.26$ to an observer's frame. Error bars show the $1-\sigma$ variation of $\tau$ for a given composition. During mixed H/He bursts (case III, red points) the relationship
between accretion rate and recurrence time is well described by a power law (red
line) for all metallicities. Simulations of case II (blue) and case IV (black) burning are
also plotted, whilst the shaded region indicates the time for hydrogen to be
depleted by the hot CNO cycle. The transition to case IV burning is expected to
occur within this shaded region. For metallicities of $\mathrm{Z}\le0.002$ we do not fit
for $\mathrm{\dot{M}}<1.4\times 10^{-9}\,\mathrm{M_\odot}\,\mathrm{yr^{-1}}$ as
bursts show an extended tail which prolongs $\Delta t$. Note that we exclude
bursts with a recurrence time less than $\Delta t = 100\,\mathrm{s}$ as the
burst duration approaches the recurrence time. This does occur at high accretion
in the $\mathrm{Z}=0.10$ and $\mathrm{Z}=0.20$} cases. Each plot has an identical power law with a
gradient of $\eta=1.0$ plotted for comparison.
\label{fig:accDel}
\end{figure*}

The relationship between accretion rate and recurrence time has been well
studied for a number of sources, including \GS\ 
\citep{Galloway2004,Thompson2008recurrenceTime} and \MXB, also known as the rapid
burster \citep{Bagnoli2013SlowRotator}. Observationally, a power law relationship
has frequently been found, relating the persistent flux, ($f_\mathrm{pers})$ of the
source to the burst recurrence time, with the form
\begin{equation}
\Delta t \propto f_{\mathrm{pers}}^{-\eta}.
\end{equation}
The persistent flux of a neutron star is primarily due to an energy deposition
of $Q_\mathrm{grav} \mathrm{g}^{-1}$ from accretion onto the star. For a star of
radius $R$ at a distance $d$ the flux (given an anisotropy of
$\xi_\mathrm{pers}$) is given by
\begin{equation}
f_{\mathrm{pers}} = \frac{L_{\mathrm{pers}}}{4\pi d^{2}} = \frac{\dot{m} Q_{\mathrm{grav}}}{1+z} \left(\frac{R}{d}\right)^{2} \xi_\mathrm{pers}^{-1},
\end{equation}
and hence ideally, the persistent flux is directly proportional to the mass
accretion rate. It then follows that variations in recurrence time are due to
changes in the accretion rate,
\begin{equation}
\Delta t \propto \dot{m}^{-\eta}.
\label{eq:delt-mdot}
\end{equation}
Na\"{i}vely, we can expect the power law gradient to be $\eta\sim 1$ if we assume
that a constant critical mass $m_\mathrm{crit}$ is required for a burst, as
$\Delta t = m_\mathrm{crit}/\dot{m}$. Deviations from this relation suggest 
that the mass necessary for an instability varies with accretion.

For each accretion composition in Table \ref{param-space}, we identified model
runs that exhibit five or more bursts, and fitted power laws for the variation in
recurrence time with accretion rate  across the region of case III burning.
Recurrence times were redshifted by $z=0.26$, to be observationally equivalent
to a ${M}=1.4\,\mathrm{M_\odot}$, $r=11.2\,\mathrm{km}$ neutron star. We
excluded models where $\Delta t<100\,\mathrm{s}$, though this only occurs for
$Z\ge0.10$. The best fit power law was found via a $\chi^2$-minimisation, using
the variation in recurrence time throughout the model run to weight each point. At the high accretion
rate end of the power law, we stopped fitting when the recurrence time stopped
decreasing. At the low accretion rate end, we stopped fitting when the points
followed a noticeably different linear trend in log-log space (noting that the nature of the deviation changes with metallicity). For metallicities 
$z=0.020$ and above, this transition corresponds to a change in the burst behaviour, from
the long-tailed lightcurve morphology (characteristic of mixed H/He bursts) to an
Eddington-limited lightcurve caused by pure He bursts. For these runs, the power
law fit spans the entire range of case III burning. The linear relationship becomes considerably steeper in the pure He bursting region. In
the low metallicity cases ($z=0.000,\ z=0.001\,\mathrm{and}\ z=0.002$), mixed H/He
burning still continues at accretion rates below where the power law stops being
fitted, as the trend is translated vertically up for $l_\mathrm{acc}<0.08$. The
value of $\eta$ and its associated 1-$\sigma$ uncertainty from the $\chi^2$
minimisation are shown in Table \ref{tab:RecTimePLaw}, and the actual data and
fits are shown in Figure \ref{fig:accDel}, with the fitted points plotted in
red.

We find that in the {\sc KEPLER} models, the value of $\eta$ depends upon the
stellar composition and accretion rate. Using the recurrence times derived from
our analysis of the model catalog we find that the variation with accretion rate
is consistent with a power law across a wide range of accretion rates and
metallicities.

There is no significant correlation between power law index and metallicity
(Spearman $\rho=-0.48$, $p=0.187\approx 1.3\sigma$). Interestingly, all
metallicities show $\eta \gtrsim 1.1$ whilst observations do not show such steep
gradients. \cite{Galloway2004}  find $\eta = 1.05\pm0.02$ for \GS\ whilst
\cite{Bagnoli2013SlowRotator} find $\eta = 0.95\pm0.03$ for \MXB.
\cite{Linares2012Terzan5} find that the X-ray pulsar \TFIVE\ shows $\eta\approx
1$ when $L_{\mathrm{pers}} > 0.3L_\mathrm{Edd}$, and $\eta\approx 3$ when
$0.2L_\mathrm{Edd} < L_{\mathrm{pers}} < 0.3L_\mathrm{Edd}$. The assumption that
the burst anisotropy is independent of accretion rate is necessary to compare
observational and theoretical power law indices. It is likely however that there
are burst behaviours that violate this assumption.


We found that models with 1\% and 2\% metallicity come closest to the observed
$\eta$ values ($\eta \approx 1.11$). It would however be overstating
the available evidence to suggest this implies that the observational sources
discussed in the preceeding paragraph are consequently metal rich, as modelled
power laws could be systematically high. It is nevertheless notable that
metallicities around solar show a minimum $\eta$ value in the parameter range
tested, suggesting that if a large sample of power law gradients were compiled,
the lowest gradients would correspond to metal-rich population I stars.

Using theoretical burst ignition models, \cite{Galloway2004} found $\eta < 1$
for solar metallicities, with $\eta$ increasing as metallicity drops. The {\sc KEPLER}
models show noticeably higher power law indices than this, though the trend for
steeper power laws to be found at lower metallicities is replicated. Part of the
reason for our disagreement with these models is that the models used in 
\citeauthor{Galloway2004} neglect additional heating of the atmosphere by
accumulated CNO, leading to systematic differences to {\sc KEPLER}.

\begin{table}[]
\centering
\begin{tabular}{cccccc}
Z & $\eta$& n & $l_{\mathrm{acc,\,IV}}$ & $l_{\mathrm{acc,\,III}}$ & $l_{\mathrm{acc,\,II}}$\\
\hline
\hline
0.000 &    $ 1.19 \pm 0.03  $ & 10 & \nodata  & $ 0.75 $ & $ 0.92 $ \\
0.001 &    $ 1.17 \pm 0.02  $ & 21 & \nodata  & $ 0.80 $ & $ 0.92 $ \\
0.002 &    $ 1.237\pm 0.014 $ & 10 & \nodata  & $ 0.80 $ & $ 0.97 $ \\
0.004 &    $ 1.17 \pm 0.02  $ & 16 & \nodata  & $ 0.80 $ & $ 1.00 $ \\
0.010 &    $ 1.109\pm 0.006 $ & 32 & \nodata  & $ 1.10 $ & $ 1.11 $ \\
0.020 &    $ 1.106\pm 0.016 $ & 21 & $ 0.03 $ & $ 1.20 $ & $ 1.26 $ \\
0.040 &    $ 1.16 \pm 0.02  $ & 19 & $ 0.07 $ & $ 1.55 $ & $ 1.71 $ \\
0.100 &    $ 1.18 \pm 0.02  $ & 17 & $ 0.20 $ & $ 4.20 $ & $ 4.20 $ \\  
0.200 &    $ 1.16 \pm 0.02  $ &  7 & $ 0.70 $ & $ 6.50 $ & $ 7.50 $ \\ \hline
{N\&H03}&                &    & $ 0.04 $ & $ 0.15 $ & $ 0.25 $ \\ 
{F81}   &                &    & $ 0.08 $ & \nodata      & $1.0$     \\

\end{tabular}

\caption{The power law index $\eta$ for the power laws fitted to case III
burning in Figure \ref{fig:accDel} are shown here for each metallicity along
with the number of fitted points, $n$. The power laws were found following a
$\chi^2$ minimisation of equation \ref{eq:delt-mdot} with respect to the data
each data set. The accretion luminosities (in terms of $l_\mathrm{Edd}$) where
burning transitions occur are also shown for each metallicity. The transition
times of \cite{Narayan2003StabilityRegimes} are reprinted from Table \ref{tab:regimes}  with the transitions predicted by \cite{Fujimoto1981accretion} also, who do not distinguish case IV from steady burning.
.
\label{tab:RecTimePLaw}}

\end{table}

We find also that the transition accretion rates, where the burning
regime changes, increase with metallicity (Table \ref{tab:RecTimePLaw}). The transition accretion rates are defined where the burst behavior of the model changes in a qualitatively similar way to that defined by the burst cases presented by \cite{Narayan2003StabilityRegimes}. We denote
$l_\mathrm{acc,\ IV}=L_\mathrm{acc,\,IV}/L_\mathrm{Edd}$ as the accretion
luminosity (in terms of Eddington) as the lowest accretion luminosity at which we
observe case III bursts. $l_\mathrm{acc,\,III}$ denotes the accretion
luminosity at which the shortest recurrence time is observed, and
$l_\mathrm{acc,\,II}$ denotes the highest accretion rate for which we observe
bursts. {\sc KEPLER} simulations produce roughly similar transition accretion rates as
stability analyses suggest, noting that both \cite{Fujimoto1981accretion} and
\cite{Narayan2003StabilityRegimes} consider solar accretion compositions.

\cite{Fujimoto1981accretion} consider three neutron star potentials, all 
significantly different from the surface gravity adopted in {\sc KEPLER}. Of the two closest to our considered surface gravity, the first overestimates
the surface gravity by $\sim 55\%$ ($M=1.00\,\mathrm{M_\odot}$,
$r=7.49\,\mathrm{km}$), whilst the second underestimates it by $\sim35\%$ 
($M=0.476\,\mathrm{M_\odot}$, $r=8.67\,\mathrm{km}$). These analyses predict
$l_\mathrm{acc,\,IV}=0.08$ for the higher gravity, and
$l_\mathrm{acc,\,IV}=0.03$ for the lighter gravity. We find
$l_\mathrm{acc,\, IV}=0.03$ for $Z=0.02$, agreeing with the lower surface
gravity case. \cite{Narayan2003StabilityRegimes} analysed the stability of
burning on a star with only a 20\% ($M=1.40\,\mathrm{M_\odot}$,
$r=10.0\,\mathrm{km}$) higher surface gravity  than {\sc KEPLER} using a more
rigorous approach, finding $l_\mathrm{acc,\,IV}=0.04$. Whilst each study
uses different surface gravities, our value for the transition rate  from case
IV to case III burning broadly agrees with earlier stability analyses.

Despite this, the transition from case III to the case II and I regimes occurs
at noticeably higher accretion rates than predicted by
\cite{Narayan2003StabilityRegimes}. We instead find these changes occur near
$l_\mathrm{acc}\sim1.2$, rather than the $l_\mathrm{acc}\sim0.25$
predicted by \cite{Narayan2003StabilityRegimes}. This is in closer agreement to
\cite{Fujimoto1981accretion} who find that the transition to stable burning
occurs at $l_\mathrm{acc}\sim1$ regardless of surface gravity. Additionally,
case II burning occurs in general over a smaller range of accretion rates than
\cite{Narayan2003StabilityRegimes} predict, however our method of choosing when
case III burning transitions to case II arguably misses part of the case II
region. This is most noticed in the $Z=0.01$ runs as case II burning shows
oscillations in recurrence rate that increase the burst rate to faster than case
III burning. Whilst in general the minimum recurrence time provides an
approximate measurement of the case III to case II transition, burning
compositions and burst morphologies would be more robust.

The general shape of the $\dot{m}-\Delta t$ relationship outside of the region
of case III burning follows the predictions of \cite{Narayan2003StabilityRegimes},
with a turnup where overstable burning commences, and a steeper power law index
in the pure He bursting regime. In the overstable burning regime (case II) the
recurrence time increases with accretion rate as a large proportion of the
accreted fuel is being burnt during accretion. As both hydrogen and helium are
burnt quickly in this regime a longer time is required to accrete sufficient
fuel for a burst to occur. The turnup is relatively small within the {\sc KEPLER}
simulations, increasing the recurrence time by a fraction of an hour. This is in
disagreement with the stability analyses of \cite{Narayan2003StabilityRegimes}
who find that the recurrence time may increase by up to a day beyond the
shortest recurrence times they observe. Whilst these two results are quantitively
different, the upturn we observe does support qualitatively the overstable
behaviour predicted for case II bursting.

\begin{figure}
\centering
\includegraphics{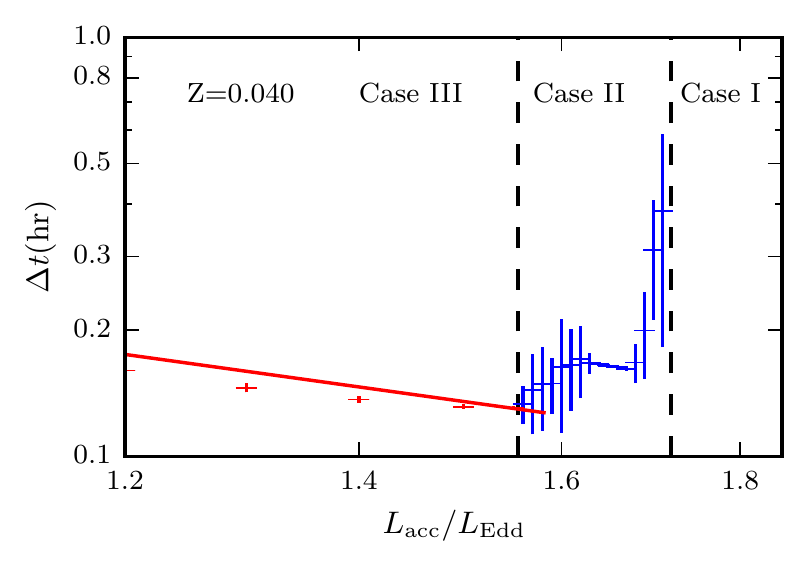}
\caption{The variation in recurrence time (plotted with $1-\sigma$ errorbars) with accretion rate for a star
accreting fuel with $z=0.040$ near the steady burning transition (case II to case I). As burning
transitions from stable H burning to overstable H/He burning, the recurrence
time abruptly begins to increase with accretion rate. The increase in recurrence
time is attributable to the increased rate of steady burning of the accumulated fuel
leading to a longer time required for a critical density to be reached.}
\label{fig:turnUpz040}
\end{figure}

The behaviour of the upturn is especially notable for the $Z=0.04$ case (Figure
\ref{fig:turnUpz040}). Whilst the upturn is reasonably brief in most
compositions, it shows substantial structure for this metallicity. Physically at
these accretion regimes the increase in recurrence time is linked to helium  in
addition to hydrogen burning between bursts depleting the available fuel. In
this regime luminosity between bursts tends not to fall quickly back to its
persistent level. Interestingly, near
$l_\mathrm{acc}\sim1.65$ the uncertainty in $\Delta t$ becomes again small,
similar to the case III uncertainties, and $\Delta t$ slightly decreases, before
resuming its upward trend. In general, the large uncertainties in $\Delta t$ in
this region are a consequence of metastable H/He burning, and the large impact
small changes in the composition of the atmosphere can have on the stability of
the accreted column. It is possible that in the intermediate area where
$l_\mathrm{acc}\sim1.65$ the column is slightly more stable, the exact reason for this will be investigated in more detail in the future

The steeper power law index in the region of pure He bursts is caused by the
higher amount of accumulated helium required for a burst as compared to
hydrogen, and the lack of heating from H burning. We could better constrain
the power law behaviour in this region with
more simulations at low accretion rates, however these typically require very
long computation times per burst, as in the helium bursting regime more
integration steps are needed per burst.


\subsection{Transitions from pure He to mixed H/He bursts}
The transition between case IV (pure He) burning and case III (mixed H/He) burning
is expected to cause a significant change in the recurrence time due to the
higher temperatures required for runaway helium burning. This leads to a
discontinuity in the $\Delta t - \dot{m}$ relationship at accretion rates low
enough for the accreted hydrogen to be depleted by hot CNO burning between
bursts.The discontinuity occurs at a recurrence time similar to the time for
the hot CNO cycle to deplete all hydrogen present. \cite{Fujimoto1981accretion}
express this relation as $\Delta t=9.7\,\mathrm{hr}\,(\mathrm{Z}/0.02)^{-1}$ in the
neutron star frame, whilst \cite{Galloway2004} suggest $\Delta
t=11\,\mathrm{hr}\,(\mathrm{H}/0.02)^{-1}(\mathrm{H}/0.7)$. We attribute the discrepancy in the \citeauthor{Galloway2004} expression due to it's adaptation of an earlier calculation in \cite{Bildsten1998Thermonuclear} that fails to incorporate neutrino energy losses in the hot CNO cycle. Given this difference however, we provide an independent derivation of this relationship
below, laying out the assumptions inherent within it.

The hot CNO cycle \citep{Mathews1984CNO}  occurs when temperatures are
sufficiently high for proton capture by $\ce{^{13}N}$ to occur more
frequently than $\ce{^{13}N}$ undergoes $\ce{\beta^+}$ decay. This leads to the hot
CNO reaction cycle
\begin{align}
\ce{^{12}C}(p,\gamma)\ce{^{13}N}(p,\gamma)\ce{^{14}O}(\ce{\beta^+}\nu)\ce{^{14}N}(p,\gamma)\nonumber\\ 
\ce{^{15}O}(\ce{\beta^+}\nu)\ce{^{15}N}(p,\alpha)\ce{^{12}C}
\end{align}
The cycle is limited by the $\ce{\beta^+}$ decay of $\ce{^{14}O}$ and $\ce{^{15}O}$. To find the time to burn all available hydrogen, we consider that each decay of $\ce{^{14}O}$ in the cycle requires four proton captures. This gives the burning rate of $\ce{^1 H}$ as follows:
\begin{equation}
\partial_{t} \mathrm{Y}_{\ce{^{1}H}}=-4\frac{\ln{2}}{t_{\frac{1}{2},\ce{^{14}O}}}\mathrm{Y}_{\ce{^{14}O}}.
\end{equation}
Integrating this expression with the initial condition that
$\mathrm{Y}_{\ce{^{1}H}(t=0)}=\mathrm{Y}_{\ce{^{1}H}}^{\mathrm{init}}$ and and finding
when the hydrogen is depleted gives the time taken to burn an accumulated fuel
layer by the hot CNO cycle,
\begin{equation}
t_\mathrm{CNO}=\frac{ \mathrm{Y}_{\ce{^{1}H}}^\mathrm{init}}{\mathrm{Y}_{\ce{^{14}O}}} \cdot \frac{t_{\frac{1}{2},\ce{^{14}O}}}{4\ln{2}}
\label{eq:t_CNO_nosub}
\end{equation}
In the hot CNO cycle, the reaction rate is restricted by the beta decay of
$\ce{^{14}O}$ and $\ce{^{15}O}$, so at any point the majority of CNO is in
these two isotopes, meaning
$\mathrm{Y}_{\ce{^{14}O}}+\mathrm{Y}_{\ce{^{15}O}}\approx \mathrm{Y}_\mathrm{CNO}$.
Noting that $Y(\ce{^{14}O})/Y(\ce{^{15}O})=t_{\frac{1}{2},\ce{^{14}O}}/t_{\frac{1}{2},\ce{^{15}O}}$ the number fraction of $Y(\ce{^{14}O})$ is expressible in terms of the CNO abundance:
\begin{equation}
\frac{\mathrm{Y}_{\ce{^{14}O}}}{\mathrm{Y}_\mathrm{CNO}}\approx \frac{ \mathrm{Y}_{\ce{^{14}O}}} {\mathrm{Y}_{\ce{^{14}O} + \mathrm{Y}_{\ce{^{15}O}}}} =\frac{t_{\frac{1}{2},\ce{^{14}O}}}{t_{\frac{1}{2},\ce{^{14}O}}+t_{\frac{1}{2},\ce{^{15}O}}}.
\end{equation}
Proceeding now from equation \ref{eq:t_CNO_nosub}, the CNO burning time can be expressed as
\begin{equation}
t_\mathrm{CNO}= \frac{\mathrm{Y}_{\ce{^1H}}}{\mathrm{Y}_{\mathrm{CNO}}} \cdot \frac{t_{\frac{1}{2},\ce{^{14}O}}+t_{\frac{1}{2},\ce{^{15}O}}}{4\ln{2}}.
\label{eq:t_CNO}
\end{equation}
Conversion of $\mathrm{Y_{CNO}}$ to a metallicity mass fraction requires
assumptions about the composition in order to find the mean molecular mass.
Assuming that all of the accreted metals are in CNO with their solar ratios
\citep{Asplund2009SolarComposition} the number fraction
$\mathrm{Y_{CNO}}=\mathrm{Z}/\bar{\mu}_\mathrm{CNO}$, with
$\bar{\mu}_\mathrm{CNO}=14.5\,\mathrm{g}\,\mathrm{mol^{-1}}$ Substituting the
half-lives of both oxygen species ($t_{\frac{1}{2},\ce{^{14}O}}=71\,\mathrm{s}$
and $t_{\frac{1}{2},\ce{^{15}O}}=122\,\mathrm{s}$) the hydrogen burning time is
given by
\begin{equation}
t_\mathrm{CNO} = 9.8\,\mathrm{hr}\left(\frac{\mathrm{H}}{0.7}\right)\left(\frac{\mathrm{Z}}{0.02}\right)^{-1},
\label{eq:delta_t_CNO}
\end{equation}
noting that the prefactor is within the precision of
\cite{Fujimoto1981accretion}, a difference that could be accounted for by a small change in the measurement of $\bar{\mu}_\mathrm{CNO}$ used.
This prefactor can show some variability depending on what the metal
component is assumed to be (noting that this variability comes from using an expression based on mass rather than number fraction). For example, assuming purely $\ce{^{16}O}$
($\ce{^{14}N}$, $\ce{^{12}C}$) leads to a prefactor of $10.8\,\mathrm{hr}$
($9.5\,\mathrm{hr}$, $8.1\,\mathrm{hr}$). Additionally, as the CNO fraction does
not typically comprise all the metallicity within a star, $9.8\,\mathrm{hr}$ is
likely an underestimate for the CNO burning time. For solar composition
\citep{Asplund2009SolarComposition}, only 66\% of the metals (by mass) are CNO.
Taking this into consideration the prefactor increases to $15\,\mathrm{hr}$ in
the neutron star frame. We do not expect the {\sc KEPLER} models to be best modelled
by this last case as the metallicity component accreted is entirely
$\ce{^{14}N}$ rather than solar.

In runs at metallicities of $\mathrm{Z}=0.02$ and higher, pure He bursts, indicated
by photospheric radius expansion, are observed at low accretion rates (blue
points at the low accretion end of the power law in Figure \ref{fig:accDel}).
For each composition, we expect the power law that describes case III burning to
have an upper bound in recurrence time for each composition to be described by
the hot CNO cycle hydrogen burning time (Equation \ref{eq:t_CNO_nosub}). For
$\mathrm{Z}=0.02\ (0.04,\ 0.10,\ 0.20)$ this time is $\Delta t = 12\ (5.7,\ 1.7,\
0.4)\,\mathrm{hr}$ after redshifting by $1+z=1.26$. In each model run  we
observe case III burning at recurrence times shorter than the hot CNO burning
time and case IV burning above these times, as expected.

This agreement is also noticed using the $11\,\mathrm{hr}$ prefactor suggested
by \cite{Galloway2004}. The modelled results in \cite{Galloway2004} have a
redshift of $1+z=1.31$, leading to a rest frame prefactor of $8.4\,\mathrm{hr}$.
Such a value could arise from accretion if most of the metallicity fraction is
carbon. Most of the suggested prefactors thus far match our simulated data.
Figure \ref{fig:accDel} shows in blue shading to highlight the range of times wherein the CNO cycle is expected to deplete available hydrogen, using the unredshifted 
prefactors of $8.1\,\mathrm{hr}$ and $15\,\mathrm{hr}$ as limits.
Only the low end ($\lesssim9\,\mathrm{hr}$) of this
range is excluded by the {\sc KEPLER} models.

Considering all this detail, we emphasise the approximate nature of equation
\ref{eq:delta_t_CNO}. Changes in accretion composition, gravitational redshift
and the presence of ashes from previous bursts can foreseeably combine to
greatly vary this prefactor. Increasing the resolution of simulations in
accretion rate in order to better isolate the transition from case IV to case
III burning may lead to a better determination of the prefactor (given the
assumptions within {\sc KEPLER}) and also allow some quantification of the deviation
CNO ashes cause from the theoretical prefactor. However as observational systems
have significantly greater deviations from theory than {\sc KEPLER}, better
constraints are unlikely to be very useful.

\subsection{Recurrence time behaviour at low metallicity}
For models run at metallicities lower than $Z=0.02$, no runs show case IV
burning, as they do not span accretion rates low enough to deplete all the
available hydrogen. Rather, a noticeable vertical translation of the power law
index occurs for accretion rates with $l_\mathrm{acc}\lesssim0.08$. To determine
the significance of this deviation we considered the $\chi^2_\nu$ for fits
to the recurrence time versus the accretion rate, across both the range of
accretion rates from $l_\mathrm{acc}=0.08$ to the case
II transition, and for all accretion rates up to the case II transition. For
$Z=0.002$ ($Z=0.001$, $Z=0.0$) we found that using the larger fitting range
drastically increased the $\chi^2_\nu$ from 0.98 (2.65,3.36) to 21.8 (33.0,
29.5). Whilst these fits are phenomenological and a high $\chi^2$ may occur on
account of this, the variation in these two fits suggests that at accretion
rates $l_\mathrm{acc}\lesssim 0.08$ the power law that describes these bursts is
different.

\begin{figure}
\centering
\includegraphics[]{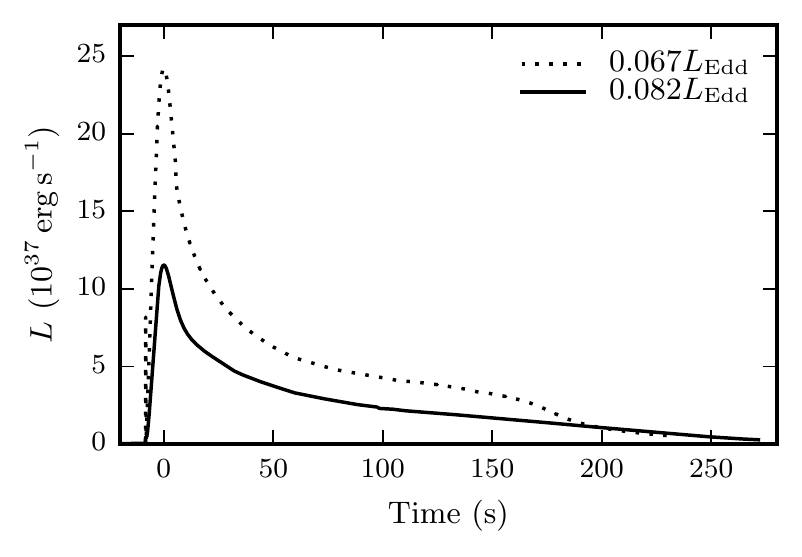}
\caption{In low metallicity sources a significant change in morphology to occurs
near $L_\mathrm{acc}\sim0.07L_\mathrm{Edd}$ ($\mathrm{\dot{M}}\sim10^{-9}\,\mathrm{M_\odot}\,\mathrm{yr^{-1}}$) whereby
the bursts become brighter and show a decay in the tail consisting of
two stages. The two models shown were run at $Z=0.001$ and have accretion rates of $L_\mathrm{acc}=0.067L_\mathrm{Edd}$ (dotted, model a20) and $L_\mathrm{acc}=0.067L_\mathrm{Edd}$ (solid, model a30). Luminosities and times are in the NS frame.
\label{fig:lowMetCompare}}

\end{figure}

To this end, we examined the morphology of bursts at this low accretion rate,
noticing that there is a change in the structure of the burst tail
at accretion rates lower than where we cease fitting. At such low accretion
rates, the burst lightcurve shows a two stage decay (Figure 
\ref{fig:lowMetCompare}) rather than a single decay. The presence of a second
decay, reached after the first decay flattens indicates an additional burning process,
which could act to consume additional fuel, thereby extending the accretion
time required to recuperate an unstable column. Further investigation into the
nuclear networks active in {\sc KEPLER} during these bursts, as well as the rapidity 
of the onset of the two stage decay, could present an interesting avenue for further study.

\subsection{Recurrence time variation with $\tau$}
\begin{figure}
\centering
\includegraphics{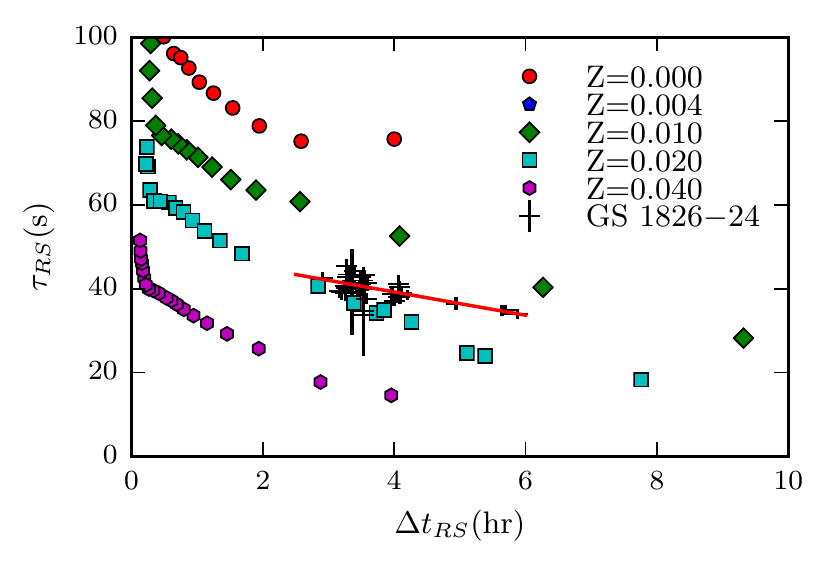}
\caption{The recurrence time and equivalent duration $\tau_{RS}$ (redshifted to the observer's frame) of a burst are observable
phenomena that vary with accretion rate. Here, accretion rate increases from right to left. Red circles,
blue pentagons, green diamonds, teal squares and purple
hexagons represent model runs with $\mathrm{Z} = 0.000,\ 0.004,\ 0.010,\ 0.020\
\mathrm{and}\ 0.040$ respectively. The black points represent observations of
the recurrence time and equivalent duration of \GS, and the associated 1-$\sigma$ error in
equivalent burst duration. The red line is a linear fit to the observed data to illustrate the general observed trend. \GS\ sits between the $\mathrm{Z=0.010}$ and $\mathrm{Z=0.020}$ models, following a shallower gradient than either.}
\label{fig:tau-tdel}

\end{figure}
Both equivalent burst duration ($\tau$) and recurrence time ($\Delta t$) are commonly used
observational measures in burst studies. Importantly, both measures are
independent of distance. As such, the loci traced by
simulated bursts in $\tau-\Delta t$ space provides a series of curves which can
be compared to observations (for an assumed gravitational redshift). This is considered in Figure \ref{fig:tau-tdel},
along with observed values of $\tau$ and $\Delta t$ for \GS. The simulated
parameters have been redshifted by $1+z=1.26$, corresponding to a
$1.4\,\mathrm{M_\odot}$ neutron star $11.2\,\mathrm{km}$ in radius.

We compared these simulated parameters to observations of \GS\ for which we had
a reliable recurrence time (determined from finding two bursts close in time).
The values of recurrence time and $\tau$ for \GS\ appear between the locus traced by the $\mathrm{Z}=0.01$ and
$\mathrm{Z}=0.02$ models. This suggests an accretion metallicity slightly lower than
$\mathrm{Z}=0.02$, in agreement with \cite{Heger2007}, although we note that their
comparison also uses {\sc KEPLER} models. The span of the observations corresponds to
modelled accretion rates of $l_\mathrm{acc} = 0.065$ to $l_\mathrm{acc}=0.123$.
To illustrate the behaviour of \GS\ we have fitted a linear model ($\tau =
A\Delta t+B$) to the observations by $\chi^2$-minimisation, finding $A =
-2.8\pm0.3\,\mathrm{s}\,\mathrm{hr^{-1}}$ and $B=50\pm1\,\mathrm{s}$
($\chi^2_\nu=1.2$).


Interestingly, the gradient from the \GS\ observations in Figure \ref{fig:tau-tdel} shows a
shallower drop in $\Delta t$ with $\tau$ than the modelled points. As higher
recurrence times are indicative of lower accretion rates, we see higher fluences
than {\sc KEPLER} predicts as accretion rate drops. The deviation from the locus of $\mathrm{Z}=0.02$ is largely driven by the three data points at recurrence times $>5\ \mathrm{hr}$. Whilst it could be tempting to to eliminate these points as examples of non-standard behavior of the star, recent observations of \GS\ have shown the star exhibiting an uncharacteristic soft spectral state \citep{Chenevez2015GSSoft}, strengthening the idea that these points reflect the behavior of \GS, and that its behavior is thus not in accordance with that predicted by {\sc KEPLER} for $Z=0.02$ accretion. Part of the reason for this may be in that the accretion regime used by us, whereby only $\ce{^{14}N}$ is being accreted, does not provide adequate seed nuclei to reproduce the behaviors of naturally occurring stars where a more diverse quantity of metals are accreted. It woud be interesting in further studies to investigate how variations in the accretion composition affect these loci.

\section{Results and Discussion II: Morphological variations}\label{sec:morphology}
A number of parameters have been used in the literature to quantify the
morphology of a lightcurve. In models, these parameters can show significantly
more variation in subsequent burst trains than is observed in some sources. In
this section we summarise the broad trends in burst convexity, $\alpha$ and tail
shape that we find in the {\sc KEPLER} models, as well as the appearance of
twin-peaked bursts.

\subsection{Burst to burst variation within a train}
A recurring puzzle in the morphologies of modelled bursts is their variablility.
Na\"{i}vely, models would be expected to show relatively uniform bursts (apart
from the first few bursts in a simulation) as the composition of the neutron
star atmosphere approaches a steady level. This is seen to some extent: the
first burst in a train is disproportionately energetic, and the following bursts
show variations of about 10\% between the peak luminosity of the brightest and
the dimmest.

This behavior implies that there is a natural variability in burst lightcurves
that arises from nuclear physics, yet some observational sources, notably \GS,
do not show as much variability within a burst epoch as simulated bursts show
within a model run. We suggest that this is due to the number of convective
cells on a bursting neutron star collectively removing the variation we see in
the models. Each model effectively assumes the neutron star contains a single
convective cell. Rather, assuming a cell diameter similar to the atmosphere
height ($\sim10\,\mathrm{cm}$), a neutron star has on the order of $10^7$
convective cells. This would produce a lightcurve equivalent to the average of
$10^7$ modelled bursts and thereby suppress the variations in burst behavior
that are caused by nuclear physics.

\subsection{Convexity}

\begin{figure}
\centering
\includegraphics{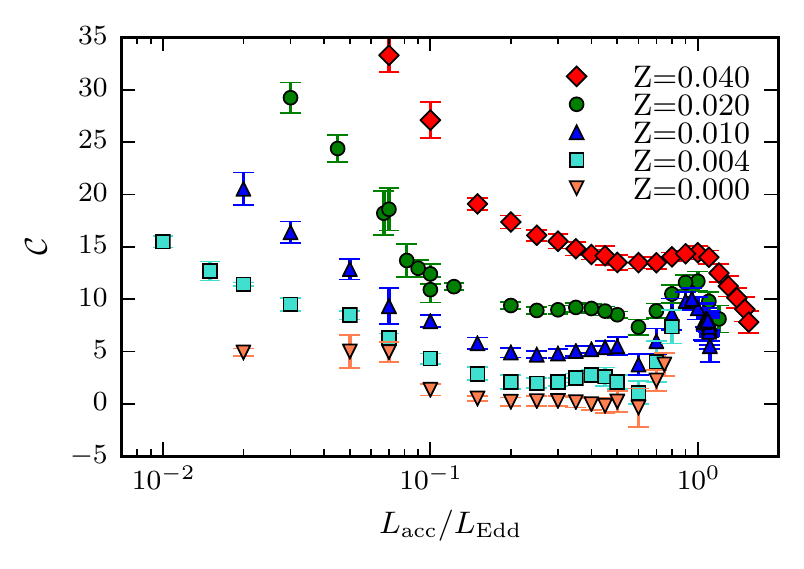}

\caption[]{Convexity increases for a given accretion rate as metallicity
increases. In general, it decrases with accretion, though it flattens in the
region $0.1\lesssim l_\mathrm{acc}\lesssim 1.0$. Here, errorbars represent the
$1\sigma$ variation in convexity within a model run. Notably, convexity shows
little variation within a model run in most cases.}
\label{fig:convexity}
\end{figure}

Variations in the shape of the burst rise can be expected as a function of the latitude
on the neutron star at which the burst commences; nuclear processes; burst spreading; and also a
contribution from scattering within the accretion disk. The {\sc KEPLER} models
simulate nuclear processes without simulating accretion disk interactions or
ignition latitude, allowing the contribution of nuclear burning to burst
convexity to be determined.

For our discussion of convexity we restrict the sample to bursts with a
metallicity of $Z\le0.04$, that exhibit a smooth rise. This excludes many PRE
bursts as these often exhibit sudden increases in luminosity from convective
zones rising to the stellar surface as burning commences. Bursts from runs with
higher metallicities were excluded as their convexities are skewed by the
development of two peaks in the rise (see \S \ref{sec:doublepeaks}).

Figure \ref{fig:convexity} shows the variations in burst convexity with both
accretion rate and metallicity. For metallicities less than $Z=0.002$, the
convexity changes very little as the accretion luminosity changes from $0.1\, 
L_\mathrm{Edd}$ to $0.6\,L_\mathrm{Edd}$. Across this range the convexity rises
with increasing metallicity, from near $\mathcal{C}=1\%$ for $Z=0.000$ to
$\mathcal{C}=9\%$ for $z=0.020$. Above $0.6\,L_\mathrm{Edd}$ bursts with 
metallicities below solar converge towards $\mathcal{C}=2\%$.

Very few `concave' bursts ($\mathcal{C}<0$) are observed in the {\sc KEPLER} sample. Rather $Z=0.0$
bursts have $\mathcal{C}=0\%$ and convexity increases with metallicity.
Variations in convexity within a run are small, with the mean uncertainty
typically on the order of one percentage point.
The burst-to-burst variation in convexity increases at high accretion rates,
correlated with the onset of case II burning, and also at low accretion rates,
where a larger amount of the accreted hydrogen is depleted.

The {\sc KEPLER} models show a far more constrained range of convexities within a run
than those observed in single sources. Following
\cite{MaurerWatts2008IgnitionLatitudeAndConvexity}, we find the mean convexity
of \UHURU \ to be $\mathcal{C}_{\mathrm{avg}} = 4.7 \%$, with a $1\sigma$ spread
of 10.8 percentage points, across a large range of persistent emissions. Even
within a restricted range of persistent flux ($2\times 10^{-9}\,\mathrm{erg}\
\mathrm{cm^{-2}}\,\mathrm{s}<f_\mathrm{pers}<3\times 10^{-9}\,\mathrm{erg}\
\mathrm{cm^{-2}}\,\mathrm{s}$) the spread in convexities is still notably larger than nuclear burning predicts. We also consider the variation in convexity
shown by \GS \ as it is known for showing consistent lightcurves dominated by
nuclear processes \citep{Heger2007}. \GS \ shows a narrower distribution of
convexities with $\mathcal{C}_{\mathrm{avg}} = 4.1\pm 2.9 \%$, though the
distribution of convexities is again wider than a single run would
on average predict. Note that the error in observational convexity values here does not consider the error in convexity from each observation as it is markedly smaller. We calculated this using the observed flux $\pm 1\sigma$, and found that for \UHURU\ and \GS, these varied on average by 0.7 and 0.3 percentage points from the convexity value calculated using the mean flux.

The larger spread in convexity for observations is in part due to the span of
persistent fluxes (and thereby accretion rates) covered by observations, as
convexity does show some variation with accretion rate. For both \GS\ and
\UHURU, this is insufficient to explain the full range in convexities spanned by
both sources. \UHURU\ shows convexities ranging from $\mathcal{C}\approx-30 \%$
to $\mathcal{C}\approx40 \%$. No single composition shows this amount of
variation in convexity across the range of accretion rates simulated, and
nuclear burning does not produce noticeably concave bursts. Hence it is likely
that physical processes (e.g. spreading) on the neutron star have a more
significant impact upon the burst rise in \UHURU\ than nuclear processes.

\GS\  shows a far more constrained range of convexities than \UHURU, however the
observed dispersion in convexities is still more than typical within the models. This
indicates that the physical processes that influence convexity are far more
constrained in the \GS\ system than in \UHURU. In their comparison of \GS\ to
{\sc KEPLER} models, \cite{Heger2007} find that \GS\ is likely accreting fuel with a
solar metallicity at $\dot{{M}}=1.58\times10^{-9}\,\mathrm{M_\odot}\
\mathrm{yr^{-1}}$. The predicted convexity from this comparison ($\mathcal{C}
\approx 11\%$) however is higher than is observed. This could indicate a
physical process on \GS\ that reduces the convexity in the rise, such as ignition
at a certain latitude and burst spreading, or it could be due to systematic uncertainties in the
{\sc KEPLER} models.

\subsection{Burst energetics}
\begin{figure}
\centering
\includegraphics{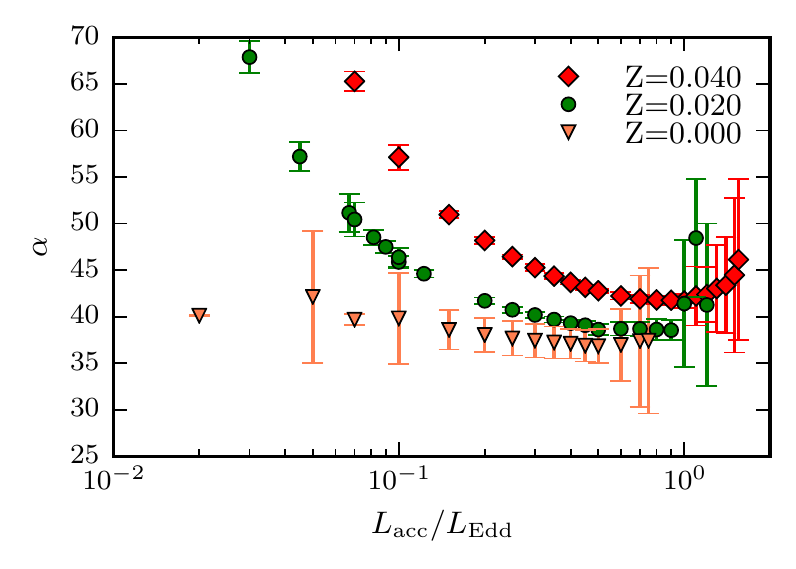}
\caption[]{The ratio of persistent fluence to burst fluence ($\alpha$) increases with metallicity. For metallicities of $Z\le0.02 $, $\alpha$ remains mostly constant across case II burning, showing an increase as the hydrogen depletes. Near metastable burning the value of $\alpha$ becomes less consistent as the burst fluence becomes more variable, and the recurrence time approaches a similar magnitude to the burst duration.}

\label{fig:alpha}
\end{figure}
We find $\alpha$ predicted by {\sc KEPLER} models match the range of $\alpha$ seen
observationally, with $\alpha \approx 40 $ in the region $0.1\lesssim
l_\mathrm{acc} \lesssim 1.0$. In general, $\alpha$ tends to grow with
metallicity (Figure \ref{fig:alpha}). At high accretion rates, variations in
$\alpha$ between bursts obscure this trend, as fluence measurements become more
variable when the recurrence time is of a similar order of magnitude to the
burst duration (typically 10\% to 50\% of $t_b$). The variation in $\alpha$ with
accretion rate matches the expected trends for each metallicity, being
approximately flat for no metallicity and increasing in curvature with
metallicity following \cite{Galloway2004}.
For the data points with zero metallicity, the uncertainties at low accretion rates are sometimes very large, this is a manifestation of large variations in peak luminosity and recurrence time between bursts, which is accentuated in the cases where large uncertainties are seen.

\subsection{Tail morphology}
\begin{figure}
\centering
\includegraphics{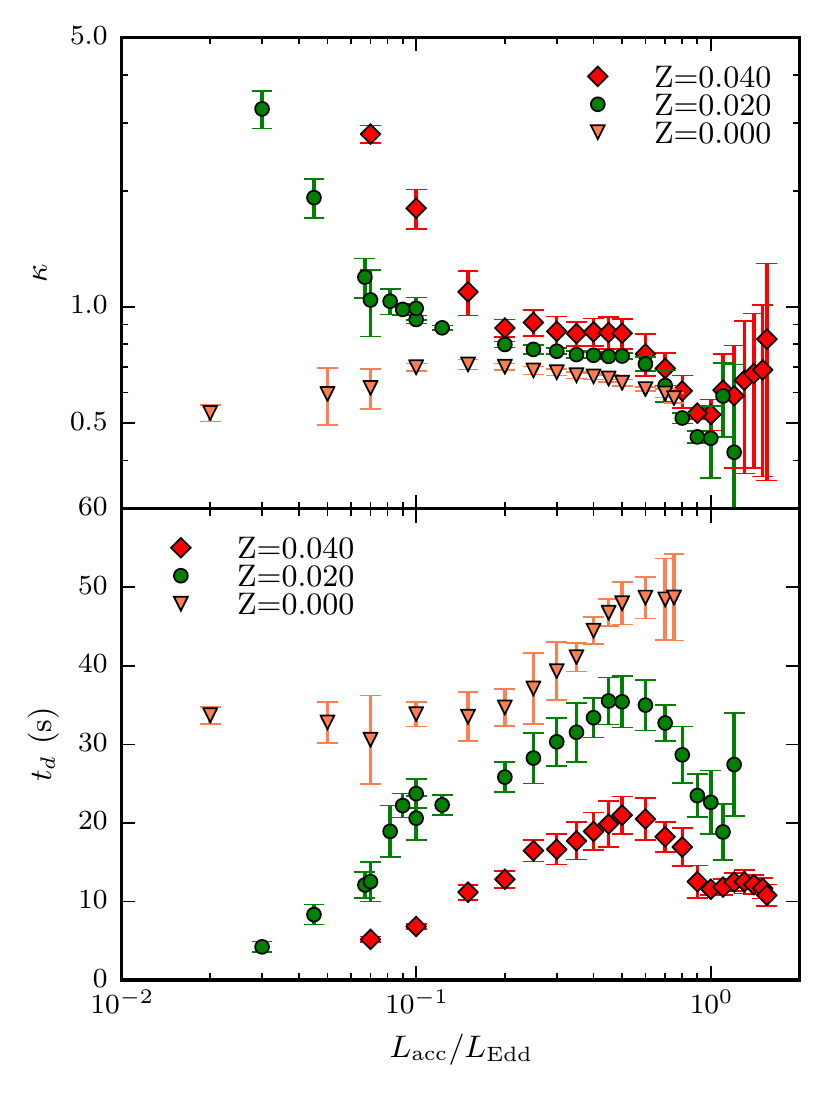}
\caption[]{As accretion rate increases, the power law index $\kappa$ of a power
law fitted to the burst tail decreases, whilst the exponential decay time of the
tail, $t_d$ increases. At higher metallicities the power law index increases at
low accretion rates, whilst the decrease at low accretion rates is a consequence
of increasingly poor fits due to a two-stage decay in the lightcurve.}

\label{fig:tails}
\end{figure}
The tails of thermonuclear bursts can yield information about the cooling
processes within the star, as well as providing a metric to enable comparisons
with observations. We found that the power law index that best describes a burst
tail increases dramatically at low accretion rates, as hydrogen is depleted and
helium becomes more prevalent in a burst. This is quite noticeable at $Z=2\%$
and $Z=4\%$ (Figure \ref{fig:tails}). In PRE bursts, similiarly high power laws
were found. If the heat in the burst is lost by radiation and the specific heat
capacity is independent of temperature, the expected power law gradient is
$\kappa=4/3$ \citep{Intzand2014CoolingTails}, although an extended $rp$-process in
the burst tail will obscure this cooling. This behaviour is especially
noticeable in the zero metallicity case, where artificially low power laws are caused
by a two stage decay (see Figure \ref{fig:lowMetCompare}).  At low accretion
rates where extended $rp$-process burning is minimal, power law gradients
approach $\kappa = 3$ and can be as high as $\kappa=6$ for some PRE bursts. Such
high power laws can be caused by degenerate photons and electrons influencing
the specific heat capacity of the atmosphere at high energies.

Across most of the range of case III burning, little variation in $\kappa$
is observed with accretion rate(Figure \ref{fig:tails}), whilst the burst decay time
increases. This is likely a consequence of the self-similarity of power laws
with scaling. The trend we see suggests increased accretion leads to longer
bursts with a similar decay power-law.

For metallicities of $Z=0.02$ and higher, the uncertainty in the power law
constant grows considerably. This occurs as a result of the burst morphology
showing large burst to burst variations as the burst duration grows similar to the recurrence
time. In most morphological parameters, the measurements tend to show large
burst to burst variations in the parameter value once the burst duration exceeds
$\approx 10\%$ of the recurrence time.

\subsection{Double-peaked bursts} \label{sec:doublepeaks}

Using the {\sc KEPLER} simulations the nuclear origins of twin peak bursts can be
investigated. As noted in simulations by \cite{Fisker2004WaitingPoints},
where two peaks occur, the first peak (which we refer to as the helium peak)
arises from the rapid combustion of a helium layer at the bottom of the neutron
star atmosphere, which convects rapidly to the surface, causing a large increase
in surface brightness. The second peak (hereafter the primary peak) occurs due
to a stalling of the $rp\mathrm{-process}$ at a nuclear waiting point. Notably,
when there is insufficient helium accumulated deep in the neutron star
atmosphere the peaks merge to create a typical mixed H/He burst.

\begin{figure}
\centering
\includegraphics{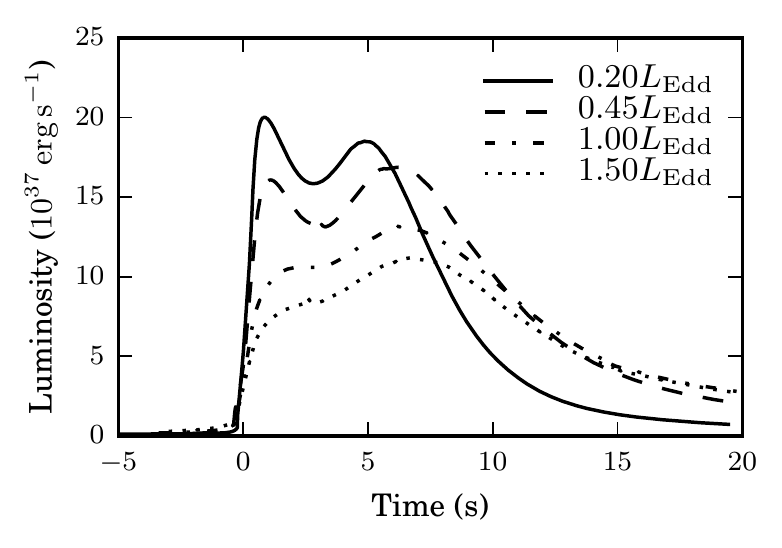}
\caption{Double peaked bursts show a dependance upon accretion rate, with the
peak becoming more pronouned at higher accretion rates. The lowest accretion
rates have the highest peak luminosity, with the solid, dashed, dot-dashed and
dotted lines show the 11th burst from models with accretion rates of
$L_\mathrm{acc}/L_{Edd}=0.20,\ 0.45,\ 1.00,\ \mathrm{and}\ 1.5$, accreting material with $z=0.10$.}
\label{fig:risesZ10}
\end{figure}

\begin{figure}
\centering
\includegraphics{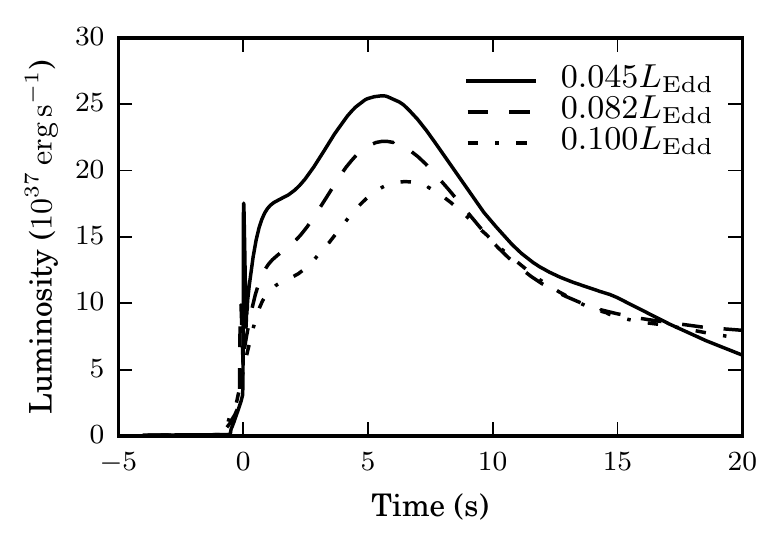}
\caption{At lower metallicities the double peak of Figure \ref{fig:risesZ10} is
not observed, but appears as a kink in the burst rise that grows more dominant
at lower accretion rates. Again, lower accretion rate models have higher peak
luminosities. The solid, dashed and dot-dashed show the 11th burst from models
with accretion rates of $L_\mathrm{acc}/L_{Edd}=0.045,\ 0.082\ \mathrm{and}\ 0.1$. The tendency of low accretion models to
exhibit rapid increases in surface luminosity from hot convective regions
rapidly rising to the surface is seen in the rise of the two lower accretion
bursts.}
\label{fig:risesZ02}
\end{figure}

In the {\sc KEPLER} models, twin peaked structures are prominent in the higher
metallicity models ($z=0.10$) at low accretion rates (Figure
\ref{fig:risesZ10}), whilst the helium peak is suppressed at higher accretion
rates, manifesting as a distinct two stage rise. Of the two
peaks present at $z=0.10$ the helium peak is larger than the primary peak at low
$\dot{m}$, and also rises faster ($\sim1\,\mathrm{s}$) than it does at higher
accretion rates ($\sim3\,\mathrm{s}$), bearing a slight resemblance to the quick
rise of a helium flash. At solar metallicities the helium peak is not present as
a local maximum but manifests rather as a kink in the burst rise (Figure
\ref{fig:risesZ02}) that is suppressed as accretion rate increases. In the
intermediate case ($z=0.04$) the helium peak is present at low accretion rates,
though the models tranisition to pure He bursts before the helium peak grows to
be equal in magnitude to the primary peak. For each case discussed, the burst
train throughout the models consistently show similar structures (twin-peaked or
kinked) beyond the first few bursts in each train, which is markedly different 
to observations, where multi-peaked bursts rarely occur regularly. We have
plotted the 11th burst for each simulation in Figures \ref{fig:risesZ10} and
\ref{fig:risesZ02} to ensure that a consistent ashes layer has been found.

The presence of twin peaked structures in our models indicates that double peaked
structures can have a nuclear origin, however a larger than typical
metallicity is required to produce these bursts consistently. The magnitude of
the helium peak is largely dependant upon accretion rate. Lower accretion rates
cause longer periods between bursts, depleting more hydrogen by hot CNO burning.
The higher relative fraction of helium, especially at the base, leads to a more
dominant initial helium powered rise.

The potential observability of a twin peaked burst however is not guaranteed as
spreading of the lightcurve in time due to the propogation of the nuclear
burning front may blur the peaks. In observational sources that show repeated
peaks, multiple peaks are not observed with regularity, possibly due to their
obfuscation by the propagating burning front. Also possible is that most
observational sources do not accrete sufficiently metallic material to display a
pronounced helium peak, however sometimes burning increases the local
metallicity enough to cause occasional double peaked bursts as a consequence
of compositional inertia.

\section{Conclusion}
In this study, we analysed the variations accretion rate and metallicity have upon measurable burst parameters with simulations run using {\sc KEPLER}. We find the following conclusions:
\begin{enumerate}
\item Within the region of case III burning, recurrence time varies with accretion rate following a power law that varies with the metallicity of the accretion source.
\item The transition between pure He and mixed H/He bursts occurs at recurrence times consistent with hydrogen depletion via the $\beta$ limited CNO cycle.
\item Modelled parameters can be compared to observational values to ascertain accretion rate and composition.
\item Nuclear burning has an effect on burst convexity, however the variation seen observationally in the burst rise is greater than nuclear burning alone predicts, supporting its origin in burst location.
\item Nuclear burning can cause two peaks to arise in thermonuclear burst lightcurves when a significant amount of hydrogen has been depleted.
\end{enumerate}

\acknowledgements
We thank Andrew Cumming for useful discussions over the years on
setting up this grid of models. NL would like to thank AH and DKG for
their guidance and support in preparing this work, and the support of
both Monash University and the Conseil General de l'Allier. AH acknowledges
support by an ARC Future Fellowship (FT120100363) and from Monash University through a
Larkins Fellowship. DKG acknowledges support by an ARC Future Fellowship (FT-0991598).
We acknowledge the support of the International
Space Sciences Institute through its support of the International Team
on Thermonuclear X-ray Bursts. This paper uses preliminary results from the 
Multi-INstrument Burst ARchive (MINBAR) which is supported under the Australian
Academy of Science's Scientific Visits to Europe program, and the Australian 
Research Council's Discovery Projects and Future Fellowships Funding Schemes.

\appendix
\section{Burst Analysis Flags}\label{app:AnalyisisFlags}
For each model run, the averaged parameters are presented in Table
\ref{tab:database}. Due to the variety of models analysed, any analysis issues
are flagged using a binary code. The flags used are: 0, indicating no analysis
issues; 1, indicating that a burst occurred as the simulation was finishing, and
as such was not analysed; 2, indicating that due to convective shocks, the
luminosity exceeded $10^{39}\,\mathrm{erg}\,\mathrm{s^{-1}}$ in the raw data,
and the lightcurve was rebinned; 4, indicating that the end of the burst was
triggered by reaching a local minimum; 8, indicating twin peaked bursts occur in
the train; 16, indicating rapid bursts occur with recurrence time less than
$100\,\mathrm{s}$; and 32, which indicates the raw data was too irregular for
analysis.

The observable effect on burst parameters of flags 2 and 8 is a substantially
poorer measurement of burst rise parameters, due to either convective shocks, or
the presence of two peaks, causing ambiguity in the identification of a peak 
luminosity. Flag 4 is a consequence of metastable H/He burning extending the
duration of a burst. In this case, the tail can sometimes show small follow-up
bursts and oscillations in brightness. As these can extend the burst, as well
as make the computation of an average burst difficult, they trigger the end of
the burst. This however causes some of the burst fluence to be excluded from 
the meausrement, and as such, fluence parameters from models flagged 4 should
be treated with caution, as they are likely to be slightly underestimated.
The analysis for rapid bursts (flag 16) is quite robust, but there
is a likelihood of a burst being missed. Additionally, these bursts arise from
quite unlikely accretion compositions ($Z>0.10$).

\section{Comparison to observations}\label{app:redshifts}
Here we briefly detail the conversion adopted used for comparisons to
observations. These allow the models to be compared to a
$1.4\,\mathrm{M_\odot}$ neutron star with a radius of $11.2\,\mathrm{km}$, 
within reasonable observational bounds for neutron star 
dimensions \citep{Lattimer2006Review}. The gravitational potential within a
neutron star
atmosphere is sufficiently intense that general relativistic effects become
relevant. As the neutron star atmosphere is small compared to the stellar
radius, Newtonian gravity approximates general relativity well within the domain
of the simulations, and all values provided in the database presented herein are
in the Newtonian neutron star frame. Due to this approximation the lightcurves
and parameters we provide require redshifting to be compared with observations.

A full derivation of all the possible scalings that can
occur given a choice of stellar radius or mass is provided in the appendix of
\cite{Keek2011Superbursts}. The gravitational potential of our simulations is
equivalent to the Newtonian potential that arises from body with a radius of $r
= 10\,\mathrm{km}$ and ${M} = 1.4\,\mathrm{M_\odot}$ (note that
quantities calculated in the Newtonian frame are unsubscripted). There is a
plurality of masses ($M_{\mathrm{GR}}$) and radii ($r_{\mathrm{GR}}$) that could
give rise to the same potential using general relativity. The locus of these
points occurs where the Newtonian and GR potentials are equal:
\begin{equation}
	GM/r^2=GM_{\mathrm{GR}}/\left(r_\mathrm{GR}^2\sqrt{1-2GM_{\mathrm{GR}}/(c^2r_\mathrm{GR})}\right).
\end{equation}
We have conducted our analysis for a neutron star
that has the same mass as the star in the Newtonian simulations, choosing $M_{\mathrm{GR}} =
{M} = 1.4\,\mathrm{M_\odot}$.  Given this choice, the radius of the
neutron star that give the same general relativistic potential increases, yielding
$r_\mathrm{GR} = \xi r=11.2\,\mathrm{km}$, where $\xi=1.12$ (note that this is not the radius measured by a distant
observer). The redshift can now be found using
\begin{equation}
	1+z = 1/\sqrt{1-2GM_{\mathrm{GR}}/(c^2 r_\mathrm{GR})} = 1.258.
\end{equation}
The relevant conversions for radius, luminosity, accretion rate, and time in 
this case for an observer (subscripted $\infty$), in terms of the Newtonian reference frame values, are:
\begin{eqnarray}
	r_\infty & = & \xi (1+z)r, \nonumber \\
	L_\infty & = & \xi^2 \cdot L/(1+z)^2,  \nonumber \\
	t_\infty & = &(1+z)t, \nonumber \\ 
	\dot{{M}}_\infty & = & \dot{{M}}, \nonumber \\ 
	\frac{L_{\mathrm{acc},\infty}}{L_{\mathrm{Edd},\infty}} & = & \frac{z}{\zeta(1+z)} \frac{L_{\mathrm{acc}}}{L_{\mathrm{Edd}}}, \nonumber \\
	\alpha_\infty & = &\frac{z}{\zeta(1+z)} \alpha ,
\end{eqnarray}
where $\zeta = G{M}/(c^2 r)$. A consequence of setting the GR mass equal
to the Newtownian mass is that the mass accretion rate is identical in the
Newtonian and observer frames. Additionally, this also yields $\xi =
\sqrt{1+z}$, which simplifies a number of these relationships. The variation in
$L_\mathrm{acc}/L_\mathrm{Edd}$ is attributable to the increased Eddington
luminosity that arises from the strengthened GR potential.

It is also worth commenting that the scaling for $\alpha$ assumes only a
contribution to the persistent emission from accretion. The radiation from
persistent emission is redshifted in the same manner as burst luminosity. This
can introduce a sytematic uncertainty into the calculation of a redshifted
$\alpha$ if the accretion component is not calculated separately from the
thermal component. As $L_\mathrm{acc}\gg L_{th}$ this is minimal, especially
given that  ${z}/{[\zeta(1+z)]} = 0.9934$ given our choice of mass.

\bibliographystyle{apj} \bibliography{references}

\clearpage
\begin{turnpage}
\begin{table*}[p]

\begin{tabular}{cccccccccccccccccc}
\centering
ID & Bursts & $Z$ & $H$ & $L_\mathrm{acc}/L_\mathrm{Edd}$ & $t_b$ & $L_p$ & $L_{th}$ & $E_b$ & $\tau$ & $\Delta t$ & $\mathcal{C}$ & $t_{10-90}$ & $t_{25-90}$ & $\kappa$ & $t_d$ & $\alpha$ & Flag \\
\nodata & \nodata & $(10^{-9}\,\mathrm{M_\odot}\,\mathrm{yr^{-1}})$ & $(\%$) & $(\%$) & $(\mathrm{s})$ & $(10^{37}\,\mathrm{erg}\,\mathrm{s^{-1}})$ & $(10^{35}\,\mathrm{erg}\,\mathrm{s^{-1}})$ & $(10^{39}\,\mathrm{erg}\,\mathrm{s^{-1}})$ & $(\mathrm{s})$ & $(\mathrm{hr})$ &  $(\%$) & $(\mathrm{s})$ & $(\mathrm{s})$ & \nodata & $(\mathrm{s})$  & \nodata & \nodata \\

\hline
\hline
a003 & 15 &  0.1  & 75.90  &  0.100  &  268.1  & 10.33   &  2.73   &  5.99   &  58.2   &   3.13  &  1.76   & 5.304   & 4.304   & 0.71    & 33.43   &   39.5  &  0 \\ 
a004 &  4 &  0.1  & 75.90  &  0.020  &  213.1  & 29.50   &  0.70   & 19.10   &  64.8   &  53.97  &  7.82   & 5.570   & 4.819   & 0.72    & 35.08   &   43.2  &  0 \\ 
a005 & 12 &  2.0  & 70.48  &  0.100  &  106.3  & 15.20   &  4.55   &  4.41   &  29.1   &   2.69  & 12.41   & 5.027   & 4.520   & 0.93    & 23.74   &   45.9  &  0 \\ 
a006 &  8 &  2.0  & 70.48  &  0.020  &   54.8  & 52.18   &  1.26   &  8.04   &  15.5   & 108.27  & 38.60   & 0.360   & 0.347   & 5.82    &  3.19   &  196.1  &  2 \\ 
a007 & 14 &  0.1  & 75.90  &  0.150  &  286.5  & 10.26   &  2.84   &  6.25   &  61.0   &   2.13  &  0.86   & 5.340   & 4.295   & 0.70    & 34.05   &   38.5  &  0 \\ 
a008 & 21 &  0.1  & 75.90  &  0.500  &  367.9  &  7.04   &  4.98   &  5.44   &  77.3   &   0.53  &  0.64   & 5.866   & 4.683   & 0.65    & 47.50   &   36.5  &  0 \\ 
a009 & 15 &  2.0  & 70.48  &  0.500  &  216.0  &  9.08   &  5.61   &  4.37   &  48.1   &   0.45  &  8.52   & 5.369   & 4.641   & 0.75    & 35.43   &   38.6  &  0 \\ 
a010 &  6 &  0.0  & 76.00  &  0.020  &  337.1  & 21.49   &  0.55   & 18.94   &  88.1   &  51.49  &  4.93   & 5.475   & 4.454   & 0.53    & 33.71   &   40.1  &  0 \\ 
a011 & 13 &  2.0  & 70.48  &  0.100  &  100.5  & 17.77   &  4.17   &  4.81   &  27.2   &   2.96  & 10.91   & 5.041   & 4.507   & 0.99    & 20.61   &   46.4  &  0 \\ 
a012 &  3 &  2.0  & 70.48  &  0.015  &   71.4  & 49.18   &  0.73   & 10.12   &  20.6   & 131.09  &  0.00   & 0.064   & 0.052   & 4.59    &  4.53   &  149.0  &  2 \\ 
a013 &  0 &  2.0  & 70.48  &  0.010  & \nodata & \nodata & \nodata & \nodata & \nodata & \nodata & \nodata & \nodata & \nodata & \nodata & \nodata & \nodata &  2 \\ 
a014 & 10 &  2.0  & 70.48  &  0.030  &   36.9  & 25.73   &  3.40   &  2.92   &  11.3   &   8.48  & 29.23   & 3.501   & 3.370   & 3.26    &  4.23   &   67.9  &  2 \\ 
a015 &  2 &  2.0  & 70.48  &  0.025  &   39.6  & 51.66   &  1.86   &  5.82   &  11.3   &  59.16  & 37.42   & 1.421   & 1.372   & 6.56    &  2.09   &  188.3  &  2 \\ 
a016 &  1 &  4.0  & 65.00  &  0.020  &   77.3  & 47.25   &  0.65   & 10.61   &  22.5   & \nodata & 47.08   & 0.419   & 0.403   & 3.67    &  5.46   & \nodata &  2 \\ 
a017 &  1 &  4.0  & 65.00  &  0.040  &   26.7  & 49.85   &  1.70   &  4.24   &   8.5   & \nodata &  0.00   & 0.076   & 0.064   & 7.64    &  1.66   & \nodata &  2 \\ 
a018 &  1 &  4.0  & 65.00  &  0.050  &   27.0  & 49.50   &  4.37   &  3.75   &   7.6   & \nodata & 34.79   & 1.600   & 1.568   & 4.74    &  1.26   & \nodata &  2 \\ 
a019 & 19 &  2.0  & 70.48  &  0.067  &   71.0  & 22.50   &  4.51   &  4.25   &  19.0   &   4.28  & 18.21   & 4.325   & 4.170   & 1.19    & 12.11   &   51.2  &  2 \\ 
a020 & 12 &  0.1  & 75.90  &  0.067  &  251.9  & 22.11   &  1.31   & 12.67   &  57.4   &  10.21  &  4.03   & 5.203   & 4.398   & 0.63    & 24.25   &   39.8  &  0 \\ 
a021 &  0 &  2.0  & 70.48  & 10.000  & \nodata & \nodata & \nodata & \nodata & \nodata & \nodata & \nodata & \nodata & \nodata & \nodata & \nodata & \nodata &  1 \\ 
a022 & 31 &  2.0  & 70.48  &  1.000  &  277.6  &  5.79   & 11.11   &  3.40   &  58.6   &   0.19  & 11.71   & 5.598   & 4.935   & 0.46    & 22.63   &   41.4  &  5 \\ 

\end{tabular}
\caption{This table presents a selection of the mean model parameters for each run in the Newtonian neutron star frame. The full contents are available online. The accretion metallicity ($Z$), hydrogen fraction ($H$) and accretion rate (as $L_\mathrm{acc}/L_\mathrm{Edd}$) are specified for each model. Additional columns online include uncertainties where available for burst duration ($t_b$), peak luminosity ($L_p$), thermal persistent emission $L_{th}$, burst fluence ($E_b$), equivalent burst duration ($\tau$), recurrence time ($\Delta t$), convexity $\mathcal{C}$, 10\%$-$90\% and 25\%$-$90\%rise time ($t_{10-90}$ and $t_{25-90}$), burst decay power law index $\kappa$, burst decay timescale $t_d$ and $\alpha$ ratio. A flag is also included to indicate any analysis issues (see Appendix \ref{app:AnalyisisFlags}).}
\label{tab:database}
\end{table*}
\end{turnpage}

\end{document}